# Differences in the Brain Waves of 3D and 2.5D Motion Picture Viewers

Seok-Hee Kim and Dal-Young Kim*

Department of Optometry, Seoul Tech, Seoul 139-743, Republic of Korea

ABSTRACT

We measured brain waves of viewers watching the 2D, 2.5D, and 3D motion pictures, comparing them with one another. The relative intensity of α-frequency band of 2.5D-viewer was lower than that of 2D-viewer, while that of 3D-viewer remained with similar intensity. This result implies visual neuro-processing of the 2.5D-viewer differs from that of the 3D-viewer.

**Introduction**

The visual perception of stereoscopic display is one of the most interesting research subjects, owing to recent increasing attention to the 3D imaging technology. Due to lack of 3D contents and difficulties in making 3D motion pictures, so-called 2.5D content (that is, 3D motion picture computationally converted from 2D one) is considered as an alternative to the real 3D contents (that is, 3D motion picture taken by dual camera configuration) [1,2].

However, certain concerns about picture quality of the 2.5D motion pictures have arisen in some previous researches, in which impairment of the picture quality of 2.5D images was reported in case of animation [3,4]. Improvised 2.5D motion pictures with low picture quality may negatively affect development of the 3D industry.

In this viewpoint, we think an analysis of the visual perception of 2.5D motion pictures is meaningful. The electroencephalography (EEG) technique was adopted for our analysis, which is widely utilized to study visual perception of human brains [5]. We could find some differences in EEG signals of the 2.5D-viewer and 3D-viewer.

**Materials and Methods**

We measured intensity (electric power) of the EEG signals from ten subjects (5 males and 5 females) watching the 2D, 2.5D, and 3D motion pictures, and compared them with one another. All of the subjects were in their twenties, right-handed, and Koreans. They also had normal stereopsis and stereoacuity, but have neither any mental illness nor brain disease.

Two kinds of motion pictures offered by nVidia Company were used for this experiment. 'Nürburgring 24 Hours Race (abbreviated to NHR)' is a real 3D motion picture made by the dual camera configuration, while 'Night's Quest (abbreviated to NQ)' is an animated 2.5D motion picture that was computationally converted from 2D one. Both of them are not just stereoscopic motion pictures but can also be displayed in 2D mode with the same contents. The motion pictures were displayed to the viewers in a sequence of the 2D-NHR, the 3D-NHR, the 2D-NQ, and the 2.5D-NQ. Each motion picture was displayed for 2 minute and 15 seconds, and the EEG signals were not recorded during the first 15 second. It was because it takes time for the viewers to get used to each type of motion pictures. The EEG signals were recorded during remnant 2 minute, and 1 minute pauses were given to the viewers between watching the each type of motion picture.

The motion pictures were displayed by shutter-glasses method, on a 22-inch wide monitor (ViewSonic VX2268), with 120Hz sync rate, with 1680×1050 resolution, and driven by nVidia 3D Vision Kit (1.5.2. version). During measurement, additional disparity of the monitor was fixed to zero, so that the viewers experienced only the binocular disparity given by the motion pictures themselves.

Wearing the 3D glasses may affect the viewer's brain and EEG intensity. In order to avoid the effect of wearing 3D glasses and compare the 2D and 3D (or 2.5D) motion pictures under the same condition, the viewers were asked to keep wearing the 3D glasses even when watching the 2D motion pictures. The amount of light incident to the viewer also may affect the EEG intensity. The amount of incident light to the 3D-viewer is just half of that of the 2D-viewers due to operation of the 3D glasses. We adjusted brightness of the monitor to 80 lux for the 2D motion pictures, while to 160 lux for the 3D (or 2.5D) motion pictures, so that the total amount of incident light is the same for both 2D- and 3D-viewer. In addition, the distance between monitor and viewers was fixed to 0.5 m for all measurements.

Electric potential of the viewer's brain was measured by a PolyG-I$^{TM}$ (Laxtha Inc.) instrument [6]. Electrodes for EEG measurements were placed at F3, F4, P3, P4, T5, T6, O1 and O2 area, according to the international 10-20 system [7]. F3 and F4 correspond to the frontal lobe, and other placements correspond to brain areas related closely with visual perception and binocular depth cells [8-11]. TeleScan $^{TM}$ program (Laxtha Inc.) was used to analyze the brain waves signals [6]. Among various frequency bands of the brain waves, we picked the α- (8~13Hz) and γ- (30~100Hz) frequency bands for analysis because they are related to relaxation of mind and (visual) perception of human brain, respectively [5]. The measured EEG intensity at each frequency band was time-averaged for two minutes, and also averaged for ten subjects. Then we calculated the relative ratio between intensity of each frequency band and that of total brain wave.

**Results**

Table 1 represents the experimental results at each electrode when the viewers watched the 2D- and 3D- NHR. Comparing the results of 3D-viewers with those of 2D-viewers, when increase or decrease was statistically significant with 95% confidence level ($p < 0.05$), we indicate corresponding p-values by asterisk. At all the electrode placements and at γ-frequency band, the relative EEG intensity of 3D-viewers tends to increase, being compared with those of the 2D-viewers. The increases at O1 and O2 were statistically significant. We could not find such increasing tendency at the α-frequency band. Increase or decrease at the α-frequency band was half to half at eight electrodes without any statistical significance.

Table 2 denotes the same experimental results when viewers watched the 2D- and 2.5D-NQ. Similarly with the results of Table 1, the relative EEG intensity of 2.5D-viewers at γ-frequency band also tends to increase, being compared with that of 2D-viewers. The increases at O1, O2, and P3 is statistically significant with 95% confidence level ($p < 0.05$), being indicated by italic bold p-values, too. On the contrary, the relative EEG intensity of 2.5D-viewers at α-frequency band is not similar with those of 3D-viewers. They tend to decrease at all the electrode placements, being compared with those of 2D-viewers. The decreases at T5, P3, P4, and O2 were statistically significant with 95% confidence level ($p < 0.05$).

**Discussion**

At the γ-frequency band, the relative EEG intensity of 3D- and 2.5D- viewers tended to increase than those of 2D-viewers. We think it is reasonable because the γ-frequency band is related to visual perception [5]. The 3D displays may require more complicate visual perception process than 2D ones, which causes more activation of visual neurons resulting in higher EEG intensity at the γ-frequency band.

Concerning the α-frequency band, it is not easy to confirm changelessness of the relative EEG intensity of 2D-and 3D-viewers, because small intensity change can be smeared due to statistical fluctuation or insufficient accuracy of measurement. With more subjects, more EEG channels, or more sensitive measurement, any weak change could be found in future study. Therefore, we are very cautious to insist that the relative EEG intensity of 3D-viewers at α-frequency band did not change from that of 2D-viewers. However, even though a future study reveals any smeared change, it seems obvious the change of relative EEG intensity of 2.5D-viewers at α-frequency band is different from that of 3D-viewers. The change of 3D-viewers is changeless or change much smaller than that of 2.5D-viewers.

It is very interesting the EEG intensity changes at the α-frequency band are dependent on whether the motion pictures are 3D or 2.5D. It is generally known the brain waves at α-frequency band are not related with visual perception, although few previous studies reported opposite results [12,13]. Due to lack of understanding of the stereopsis mechanism, we cannot fully elucidate what in the visual information processing makes the difference in EEG intensity. However, it must be emphasized the 3D and 2.5D motion pictures are differently perceived by the viewers' brain, and the computationally converted 2.5D motion pictures are not perfect countermeasures to the real 3D ones.

**Conclusions**

When viewers watched the 3D display, changes of the relative EEG intensity at α-frequency band was dependent on whether displayed motion picture was 3D or 2.5D. These results can be interpreted to difference between the 3D and 2.5D motion pictures in visual neuro-processing of brain.

**Figures**

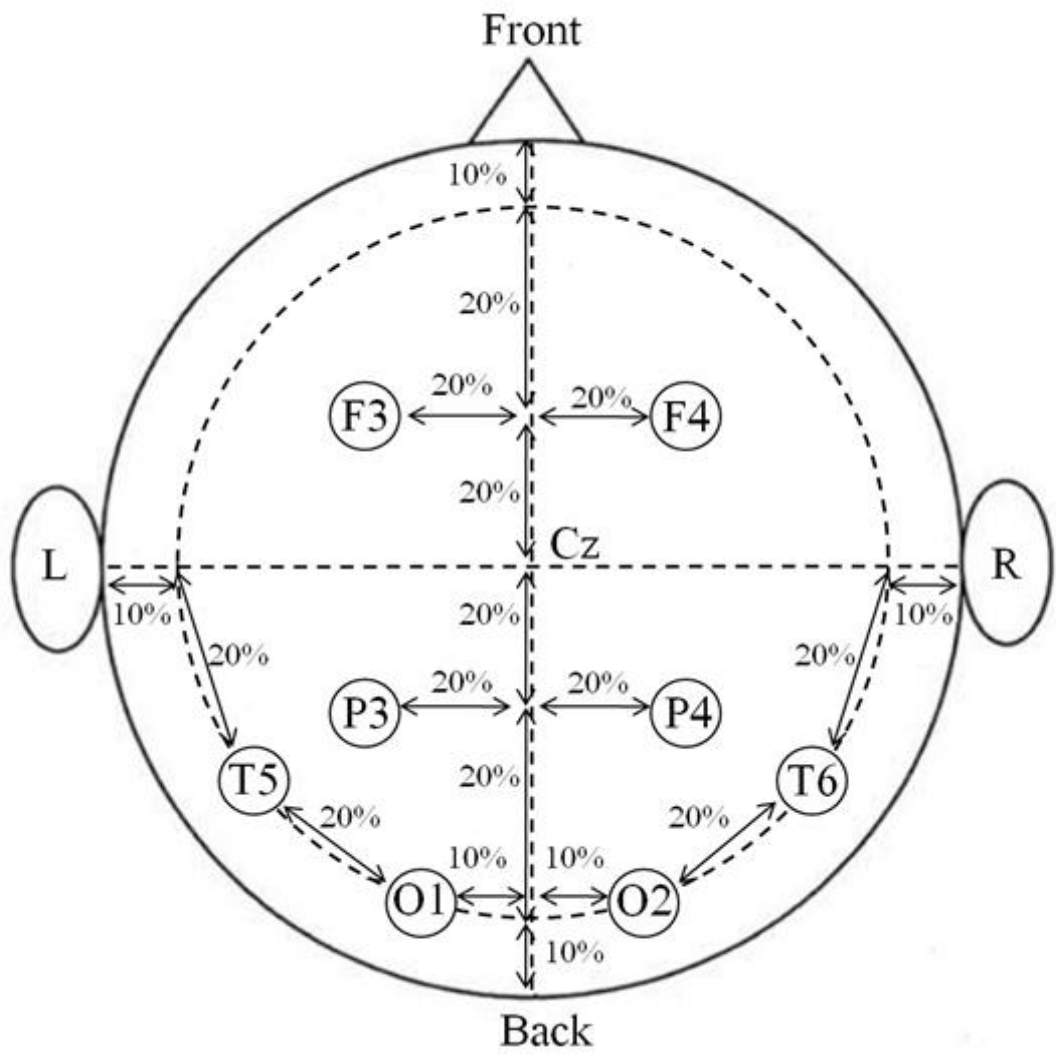

Fig. 1. Electrode locations for EEG measurement.

**Tables**

Table 1. Averaged EEG intensity at γ- and α-frequency band at each electrode placement when viewers watched real 2D and 3D motion pictures.

| Electrode placement | γ-frequency band | | p-value | α-frequency band | | p-value |
|---|---|---|---|---|---|---|
| | 2D | 3D | | 2D | 3D | |
| F3 | 0.0906265 | 0.1286769 | 0.171 | 0.1687555 | 0.1709381 | 0.857 |
| F4 | 0.1236259 | 0.1314636 | 0.690 | 0.1555622 | 0.1704152 | 0.120 |
| T5 | 0.1503934 | 0.1698375 | 0.330 | 0.2327323 | 0.2392847 | 0.776 |
| T6 | 0.2209654 | 0.2339880 | 0.442 | 0.1923086 | 0.2037445 | 0.469 |
| P3 | 0.1448089 | 0.1729071 | 0.079 | 0.2348707 | 0.2237512 | 0.497 |
| P4 | 0.1183750 | 0.1496891 | 0.125 | 0.2377980 | 0.2371167 | 0.973 |
| O1 | 0.1469305 | 0.2167686 | 0.028* | 0.2215192 | 0.2003209 | 0.142 |
| O2 | 0.1490194 | 0.2056667 | 0.045* | 0.2181426 | 0.2033954 | 0.268 |

* $p < 0.05$

Table 2. Averaged EEG intensity at γ- and α-frequency band at each electrode placement when viewers watched 2D and 2.5D animations.

| Electrode placement | γ-frequency band | | p-value | α-frequency band | | p-value |
|---|---|---|---|---|---|---|
| | 2D | 2.5D | | 2D | 2.5D | |
| F3 | 0.1237492 | 0.1474597 | 0.289 | 0.1779545 | 0.1648330 | 0.104 |
| F4 | 0.1273105 | 0.1283555 | 0.933 | 0.1675611 | 0.1636959 | 0.129 |
| T5 | 0.1456765 | 0.1858341 | 0.061 | 0.2647398 | 0.2342004 | 0.023* |
| T6 | 0.2049010 | 0.2348952 | 0.108 | 0.2124955 | 0.1920855 | 0.088 |
| P3 | 0.1270082 | 0.1581916 | 0.026* | 0.2630606 | 0.2373578 | 0.019* |
| P4 | 0.1307470 | 0.1438878 | 0.437 | 0.2501135 | 0.2276037 | 0.025* |
| O1 | 0.1513482 | 0.1937135 | 0.047* | 0.2408872 | 0.2142739 | 0.080 |
| O2 | 0.1477896 | 0.1758493 | 0.016* | 0.2405690 | 0.2163311 | 0.018* |

* $p < 0.05$